\documentclass[aps,pra,superscriptaddress,twocolumn,longbibliography,floatfix]{revtex4-1}
\usepackage{mathrsfs}
\usepackage{amsfonts}
\usepackage{amssymb}
\usepackage{amsmath}
\usepackage{graphicx}

\graphicspath{{figures/}{./}} %Setting the graphics path
\usepackage{sidecap}
\usepackage{color}
\usepackage{xcolor}
\usepackage[colorlinks=true, urlcolor=blue, citecolor=blue,linkcolor=blue,citebordercolor={1 0 0},linkbordercolor={0 0 1}]{hyperref}
\usepackage{subeqnarray}
\usepackage{verbatim}
\usepackage{euscript} %curly fonts mathcal and mathscr

\usepackage{hyperref}

\usepackage{bibunits}

\usepackage{array}
\usepackage{multirow}

\usepackage{braket}

\usepackage{algorithm}
\usepackage{algpseudocode}

\usepackage{placeins}
\renewcommand{\eqref}[1]{Eq.~(\ref{#1})} % Reference to equation
\newcommand{\figref}[1]{Fig.~\ref{#1}} % Reference to figure
\newcommand{\miteecs}{Department of Electrical Engineering and Computer Science, Massachusetts Institute of Technology, Cambridge, MA 02139, USA}
\newcommand{\harvard}{Department of Physics, Harvard University, Cambridge, MA 02138, USA}
\newcommand{\rle}{Research Laboratory of Electronics, Massachusetts Institute of Technology, Cambridge, MA 02139, USA}
\newcommand{\mitphysics}{Department of Physics, Massachusetts Institute of Technology, Cambridge, MA 02139, USA}

\newcommand{\ju}{Department of Computational Methods in Chemistry, Jagiellonian University, Gronostajowa 2, 30-387 Kraków, Poland}
\newcommand{\zapata}{Zapata Computing, Inc., 100 Federal St, 20th Floor, Boston MA, 02110 USA}
\newcommand{\BMW}{BMW Group Information Technology Research Center, 2 Research Dr., Greenville, SC 29607, USA}

\begin{document}
    
    \title{Quantum-Inspired Optimization for Industrial Scale Problems}
    
    \author{William P. Banner}
    \thanks{These authors contributed equally to this work.}
    \affiliation{\miteecs}
    
    \author{Shima Bab Hadiashar}
    \thanks{These authors contributed equally to this work.}
    \affiliation{\zapata}
    
    \author{Grzegorz Mazur}
    \thanks{These authors contributed equally to this work.}
    \affiliation{\zapata}
    \affiliation{\ju}
    
    \author{Tim Menke}
    \thanks{Current address: Atlantic Quantum, Cambridge, MA}
    \affiliation{\rle}
    \affiliation{\mitphysics}
    \affiliation{\harvard}
    
    \author{Marcin Ziolkowski}
    \affiliation{\BMW}

    \author{Ken Kennedy}
    \thanks{Current address: Fulfilld.io, 70 South Cherry St, Denver, CO 80246}
    \affiliation{\BMW}
    
    \author{Jhonathan Romero}
    \affiliation{\zapata}

    \author{Yudong Cao}
    \email{yudong@zapatacomputing.com}
    \affiliation{\zapata}
    
    \author{Jeffrey A. Grover}
    \affiliation{\rle}

    \author{William D. Oliver}
    \email{william.oliver@mit.edu}
    \affiliation{\miteecs}
    \affiliation{\rle}
    \affiliation{\mitphysics}
    
    \date{\today}
    
    \begin{abstract}
        Model-based optimization, in concert with conventional black-box methods, can quickly solve large-scale combinatorial problems. Recently, quantum-inspired modeling schemes based on tensor networks have been developed which have the potential to better identify and represent correlations in datasets. Here, we use a quantum-inspired model-based optimization method TN-GEO to assess the efficacy of these quantum-inspired methods when applied to realistic problems. In this case, the problem of interest is the optimization of a realistic assembly line based on BMW's currently utilized manufacturing schedule. Through a  comparison of optimization techniques, we found that quantum-inspired model-based optimization, when combined with conventional black-box methods, can find lower-cost solutions in certain contexts.
    \end{abstract}
    
    \maketitle
        
    \section{Introduction}
        Large-scale integer combinatorial optimization problems represent one of the broadest classes of hard  problems relevant to real-world applications \cite{Research_trends_Weinand}. A major difficulty in executing such tasks lies in the inability of conventional optimization methods to effectively sample the full state space. In addition, the constraints of a discrete optimization space preclude the use of efficient optimization methods such as the simplex method or Bayesian Optimization method which rely on smoothness to properly estimate correlations and gradients. Commonly tested examples of combinatorial problems in the literature include the travelling salesman problem or max-cut problem. These problems have a well-defined and easily-representable structure that can be exploited during solving. In practice, however, many combinatorial problems are difficult to write so explicitly, and fall under the category of black-box problems. These problems include scheduling and design problems.
        
        The curse of dimensionality, endemic to combinatorial problems, has been greatly diminished by the combined use of conventional optimization methods with machine learning techniques \cite{cheng2018,surrogate_original,surrogate_recent_minku,Data_driven_EA_survey}. Historically, these models have taken a number of forms, but many popular methods include Bayesian optimization, Model-based Evolutionary Algorithms, and Cross Entropy Methods \cite{EGO_Jones1998,Model_based_zlochin,Cross_entropy_Rub_kro}. 
        
        In parallel with these advances, new techniques for generative modeling known as Born machines have been developed based on physical phenomena -- specifically quantum systems \cite{bornmachine_Benedetti_2019,Huggins_TNML}. In recent years, such generative models have been proposed to assist in solving combinatorial optimization problems through the Generator-Enhanced Optimization (GEO) framework \cite{alcazar2021:geo_enchancing_combinatorial}. In that work, the authors used quantum-inspired model known as Tensor Network Generator Enhanced Optimization (TN-GEO), which uses a tensor-network-born-machine-based generative model to enhance the performance of conventional optimization methods. More specifically, the authors demonstrate their framework by applying it to a realization of the portfolio optimization problem. These ``quantum-inspired" generative models include resources such as entanglement to capture dependencies within a system that are not explicitly known a priori \cite{Jav_sym_tensor_networks,TTN_classifiers_for_ML,QI_Comb_opt_Hao,QI_Genetic}. This change in representation as compared with conventional methods may provide new heuristics for exploitation during optimization.

        In this work, TN-GEO was used to assess the performance of quantum-inspired optimization in solving a particular instance of a hard industry-relevant combinatorial optimization problem, solving a BMW production planning problem. The instance that we consider is realistic, with practical utility to BMW and to other manufacturers more broadly. We directly compare the performance of several related  generator enhanced optimizers to the performance of conventional optimization methods in order to determine the existence and type of advantage gained by the model-based process.
        
        We found that quantum-inspired generator enhanced optimization either meets or exceeds conventional methods of optimization for the BMW production planning problem in a majority of tested cases. The highest performance was achieved when the problem was formulated such that the correlations between parameters were efficiently encoded. In fact, these ``problem-inspired" configurations improved the performance of all the considered optimizers. Our results support the conjecture that problem-specific knowledge is an important component of optimizer selection, even among blackbox optimization methods \cite{Domain_knowledge_2018,domain_knowledge_2021,domain_knowledge_2022}. In addition, we found that intermediate problem knowledge was best for accurate model-based optimization in this setting.
        
        In the following, we describe the BMW production planning problem formulations, explain the use of problem knowledge, in particular the preprocessing of cost-function evaluations, define the optimizers used for benchmarking and analysis, and show numerical results characterizing optimizer performance. We conclude by summarizing the best contexts for model-based optimization.
        
        % this is a good sentence, looking for a good place in the text
        % With this work, we hope to provide a guide for future best-practices when using generative-model augmented optimization.
        %%\\~\\
    
    \section{Problem description and formulations}
        %%%%%%%% FIGURE: Assembly Line Description %%%%%%%%
        \begin{figure}
        \centering
        \includegraphics[scale=0.6666]{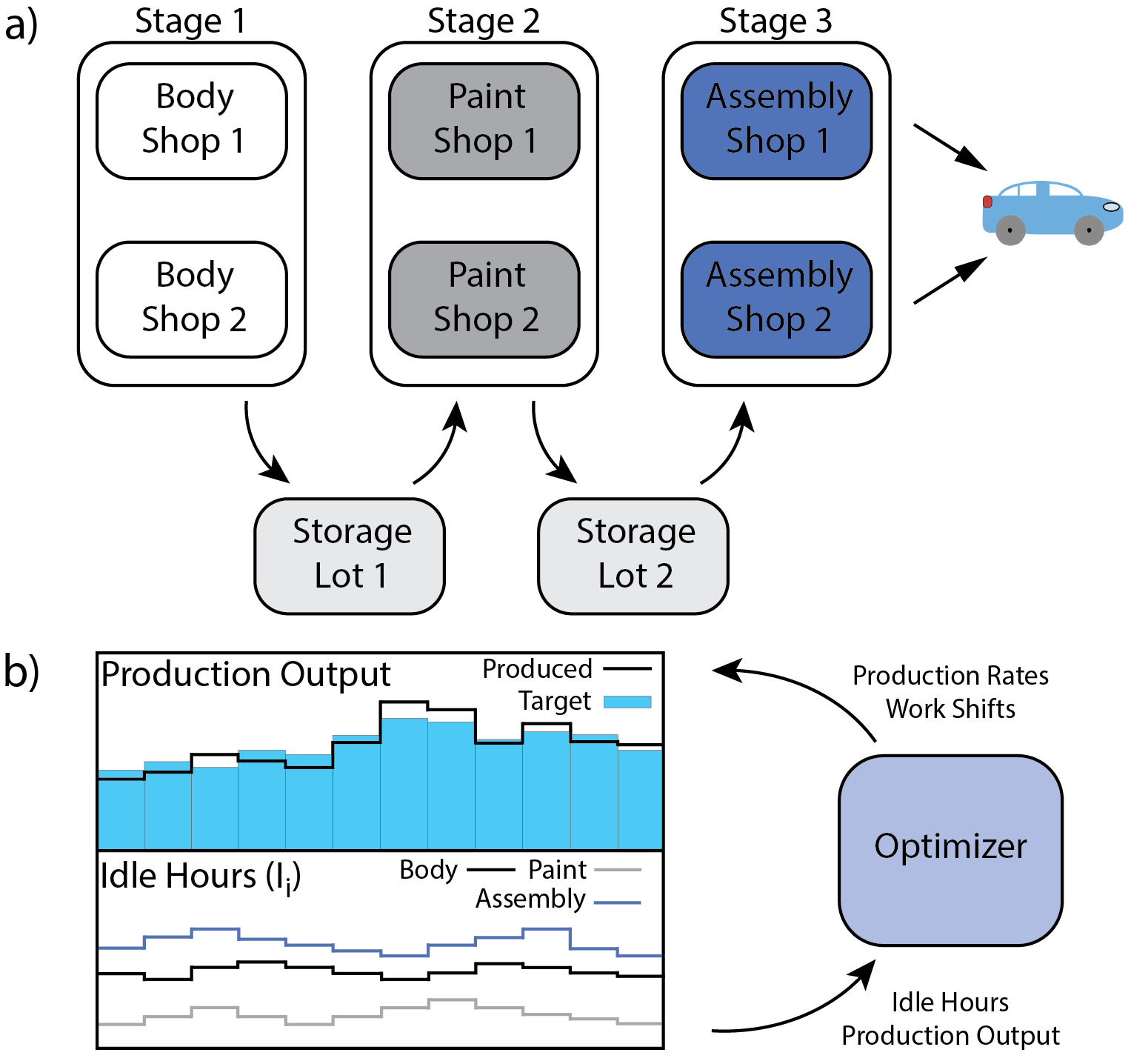}
        \caption{A) A high-level view of the production planning problem with shops and storage labeled. B) A stylized diagram of the optimization procedure. First, the parameters of the optimization, in the form of production rates and work shifts, are used to perform a time-domain simulation of the assembly line. From this simulation, the number of idle hours for each month and cars produced each month is retrieved. These values are used to generate a cost using the cost function.
        }
        \label{fig:intro}
        \end{figure}
        %%%%%%%%%%%%%%%%%%%%%%%%%%%%%%%%%%%

        The production planning problem involves the optimization of a BMW assembly line. In a BMW assembly line, as shown diagramatically in \figref{fig:intro}, car production proceeds as follows:
        \begin{enumerate}
            \item Car bodies are fabricated in a first stage consisting of two parallel body shops.
            \item Car bodies are stored in a storage lot with capacity limit 500.
            \item Car bodies are painted in a second stage consisting of two parallel paint shops.
            \item Painted car bodies are stored in a storage lot with capacity limit 700.
            \item Finished cars are assembled in a third stage consisting of two parallel assembly shops.
        \end{enumerate}
        
        The problem is constrained such that each shop operates only during specific time periods or shifts and at certain rates. There are 15 discrete options for the shifts that shops can follow and 5 discrete options for the rates. Given that there are 6 different shops, there are about $244$ million ways in which the assembly line can be configured.
        
        Over the course of one year, the assembly line is required to produce a number of cars (P$_i$) to meet production targets (T$_i$) on the order of thirty thousand cars per month $i$. At the same time, the line should be as efficient as possible, minimizing the hours per month $i$ during which each shop $j$ is idle (I$_{ij}$). This idling is due to storage lot overflows or underflows caused by mismatched shop working schedules. These two performance metrics -- production and efficiency -- are quantified using a weighted (W) cost function.
        \begin{align*}
            C &= \sum_{i = 1}^{12 \text{ months}} |\text{T}_i - \text{P}_i| + \text{W}  \sum_{\text{j}=1}^{6 \text{ shops}} \text{I}_{ij}.
            \label{eq:plant cost}
        \end{align*}
        
        For the purposes of this work, the weighting (W) was chosen such that typical variations in idle hours are on the same order as monthly production and target variations. For our specific problem, idle time with respect to scheduled time typically varies at the single percentage level, whereas production variations with respect to targets are on the order of hundreds of cars per month, leading to a weighting of 1000. For a given assembly line state, that is, the set of shift schedules and production rates used by the line, a time-domain simulation of the assembly line is performed which directly tracks the monthly production and idle hours. Additional details of the simulation are provided in Section \ref{sec:TD Sim}. An illustration of the time-domain simulation process is given in \figref{fig:intro}. 
        
        % Two different parameterizations were investigated in order to better elucidate the impact of problem knowledge on problem performance. By parameterization, we mean the manner in which the production rates and shift schedules can be changed. The first parameterization is a basic ``12-body" or ``no-knowledge" parameterization. This parameterization considers all parameters as independent with 6 shift parameters, each ranging from 1 to 15, and 6 production rate parameters, each ranging from 1 to 5. A change in shift can therefore happen in a single update step (changing a 4 to a 5, for example). This parameterization has the benefit of assuming little about inter-shop interactions, allowing solvers to explore these themselves. A second ``three-body" or ``problem-inspired" parameterization is also explored, which uses the assembly line structure to inform parameter representation. Each set of body, paint, and assembly shops are combined into a single data entry, such that the problem reduces to only three parameterized stages, with each stage having enumerated states ranging from 1 to $5625$ (the total number of possible states for one stage with two shops, each having one shift parameter with 15 options and one production rate parameter with 5 options). A change in shift may now effectively require several steps as opposed to a single step. Such a parameterization treats correlations between stages as having less impact than those within stages, an assumption based on problem intuition which may not be valid for all parameter settings. 

        Two different parameterizations were investigated to elucidate the impact of problem knowledge on problem performance. By parameterization, we refer to the manner in which production rates and shift schedules are represented and can be changed. The first parameterization is a basic ``12-body" or ``no-knowledge" parameterization. This parameterization considers all parameters as independent with 6 shift parameters, each ranging from 1 to 15, and 6 production rate parameters, each ranging from 1 to 5. A change in shift can therefore occur in a single update step (e.g., changing a shift parameter from shift schedule 4 to shift schedule 5). This parameterization is beneficial as it assumes little about inter-shop interactions, allowing solvers to explore these interactions themselves.
        
        A second ``three-body" or ``problem-inspired" parameterization is also explored, which uses the assembly line structure to inform parameter representation. Each set of body, paint, and assembly shops is combined into a single data entry, such that the problem reduces to only three parameterized stages. In this case there are \(5625\) states for every stage, accounting for the total number of possible states for two shops, each having one shift parameter with 15 options, and one production rate parameter with 5 options. The single-stage states are then ordered by their individual ideal output in cars per hour. Such a parameterization treats correlations between stages as having less impact than those within stages, an assumption based on problem intuition that may not be valid for all parameter settings.
        
    \section{Preprocessing}
    
    \subsection{Search space reduction}
        Preprocessing is utilized to effectively reduce the problem space in a deterministic manner, which is a practical and often necessary step for many large optimization problems \cite{search_space_reduce,search_space_reduce_2}. The preprocessed configurations extrapolate the ideal annual car production of each assembly line state when buffer limits are ignored and compare it to the annual production target, that is, the sum of the monthly targets (see Appendix \ref{sec:production_estimation} for details). If the state produces enough cars to be within 5\%, 2.5\%, 2\%, or 1.5\% of the annual target, it is included in the allowed state space. The preprocessed configurations are therefore referred to as ``Reduced-5\%", ``Reduced-2.5\%", ``Reduced-2\%", and ``Reduced-1.5\%" (see Appendix \ref{sec:reduction} for  details). However, we note that the minimum reduction percentage that conserves the global minimum is not known \textit{a priori}. This was verified in the cases studied here.

        \begin{table}
            \centering
            \caption{Sizes of the reduced solution spaces.}
            \label{tab:solution_space_size}
            \begin{tabular}{llr}
            \toprule
            \multicolumn{2}{c}{Solution space} & Solution space size \\
            \colrule
            2\% & noDev & \makebox[85pt][r]{384}\phantom{100} \\
            2.5\% & noDev & \makebox[85pt][r]{1056}\phantom{100} \\
            5\% & noDev & \makebox[85pt][r]{11856}\phantom{100} \\
            100\% & noDev & \makebox[85pt][r]{11390625}\phantom{100} \\
            \colrule
            1.5\% & yesDev & \makebox[85pt][r]{4777500}\phantom{100} \\
            2\% & yesDev & \makebox[85pt][r]{12329982}\phantom{100} \\
            2.5\% & yesDev & \makebox[85pt][r]{22261792}\phantom{100} \\
            5\% & yesDev & \makebox[85pt][r]{202461840}\phantom{100} \\
            100\% & yesDev & \makebox[85pt][r]{177978515625}\phantom{100} \\
            \botrule
            \end{tabular}
        \end{table}

    \subsection{State encoding}
        In the ``three-body'' representation (whether reduced or not), every state is designated by a triple of natural numbers. However, TN-GEO requires states to be represented as bitstrings (sequences of binary digits). In this work, we compare three different binary encodings. The simplest approach is to enumerate all triples, and then encode the number of every state as a binary number. We call this the basic encoding.
        
        While simple, the basic encoding may not adequately represent the structure of the problem. This is because states that are close to each other by parameterization may have significant Hamming distance incurred via the basic encoding. We tested a method of alleviating this by applying the Gray encoding~\cite{Gray_1953} to every part of the triple, padding with zeros if necessary, and concatenating into a bitstring.
    
        A significant factor that may negatively impact the effectiveness of both of the straightforward representations described above is the fact that nearest neighbours may have significantly different production costs. We test a method for minimizing this impact by enumerating the single-stage states of the first stage according to the absolute value of the difference between the estimated production of the first stage and the target production. For the remaining stages we use a slightly different ordering, employing the absolute value of the difference between estimated production of the given stage and that of the first stage. We call this reordering the Production Guided (PG) encoding (see Appendix \ref{sec:pg_encoding} for details). Then we apply the Gray encoding to every part of the triple, pad with zeros if necessary, and concatenate into a bitstring. We call this two-step encoding the PGGray encoding.
    
    \section{Optimization}
        \label{sec:methods of opt}
        %2nd draft of this section
        The conventional optimization methods in this work include one-crossover, two-crossover, and uniform-crossover genetic algorithms (GA1, GA2 and GAU, respectively), simulated annealing (SA)~\cite{KGV1983-SA,pt_swendsen}, and parallel tempering (PT) algorithms. These algorithms are the most commonly investigated algorithms in the literature and should provide an even benchmark for testing TN-GEO performance \cite{Research_trends_Weinand}. These conventional methods run until 240 cost function evaluations are reached, a limit chosen to fall within the typical black-box optimization run times of hundreds of cost function evaluations.
        
        \begin{figure*}
            \centering
            \includegraphics[width=\textwidth]{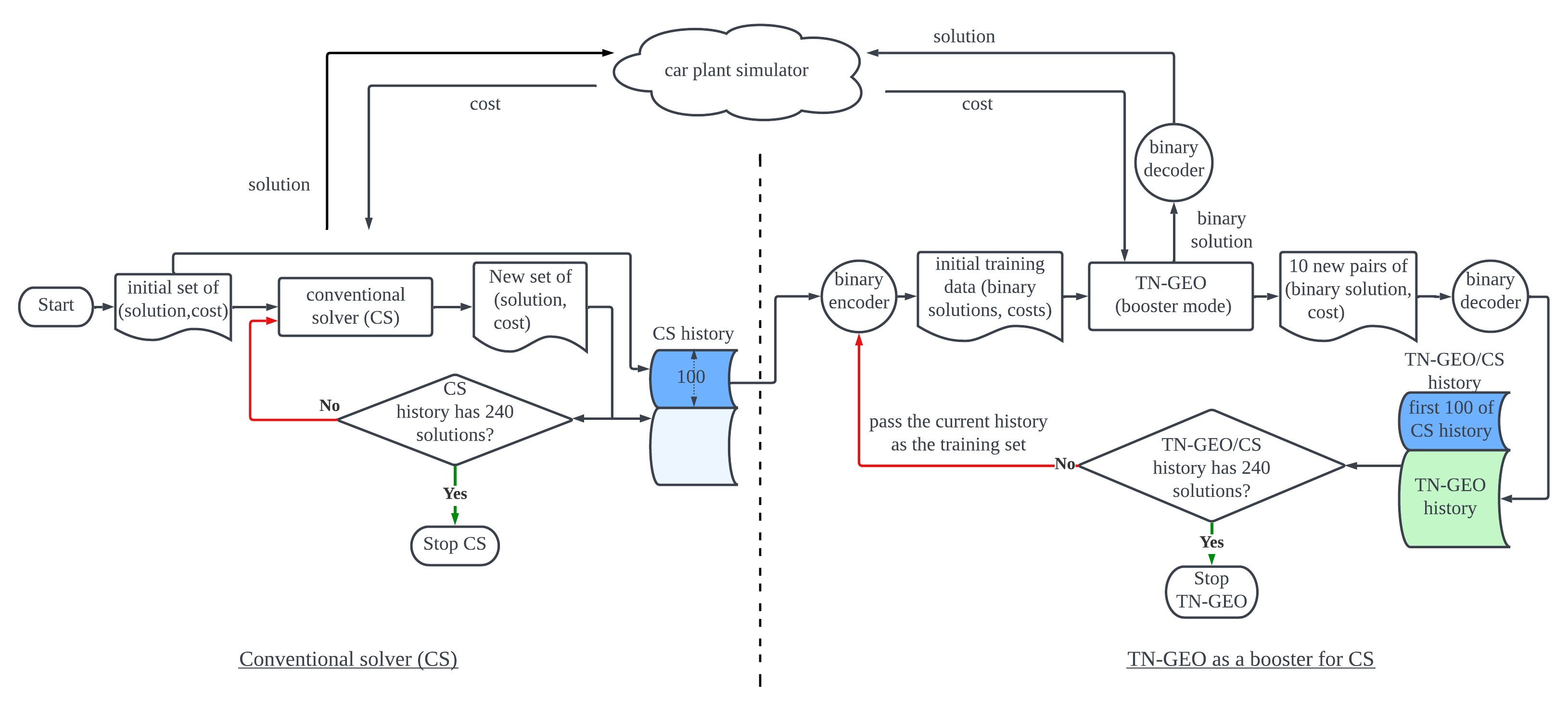}
            \caption{Scheme of our approach for benchmarking a conventional solver against TN-GEO as a booster for that conventional solver.}
        \label{fig:optimization_workflow}
        \end{figure*}
        
        When using TN-GEO to boost a conventional solver, the first 100 cost function evaluations of the conventional solver are incorporated into the TN-GEO model, and model-based optimization begins. The quantum-inspired model-based optimization step relies on TN-GEO, a method described in Ref.~\cite{alcazar2021:geo_enchancing_combinatorial} and illustrated in \figref{fig:optimization_workflow}. This model uses a Matrix Product State (MPS), a specific type of Tensor Network (TN), to represent the solution space. In each iteration of TN-GEO, the model is first trained on a seed data set of all evaluated states and their costs up to that iteration. This set is initially generated by the conventional optimization step. Training of TN-GEO on this data set is performed based on a gradient method used in a previous study \cite{mps_gen_mod}. The TN-GEO model is then sampled to generate a set of likely low-cost states. From these states, the ones that are new (not evaluated before) are then pruned according to problem constraints and a fixed-size subset of them is chosen (according to some heuristics) to have their costs evaluated explicitly on the BMW production planning problem. These states and their costs get added to the seed data set and the solver iterates until convergence or until a maximum number of cost evaluations is reached. (See Fig.~1 of Ref.~\cite{alcazar2021:geo_enchancing_combinatorial} for more details.)

    \section{Results}
        \label{sec:numerical_results}
        
\begin{figure}
        \centering
        \includegraphics[width=\columnwidth]{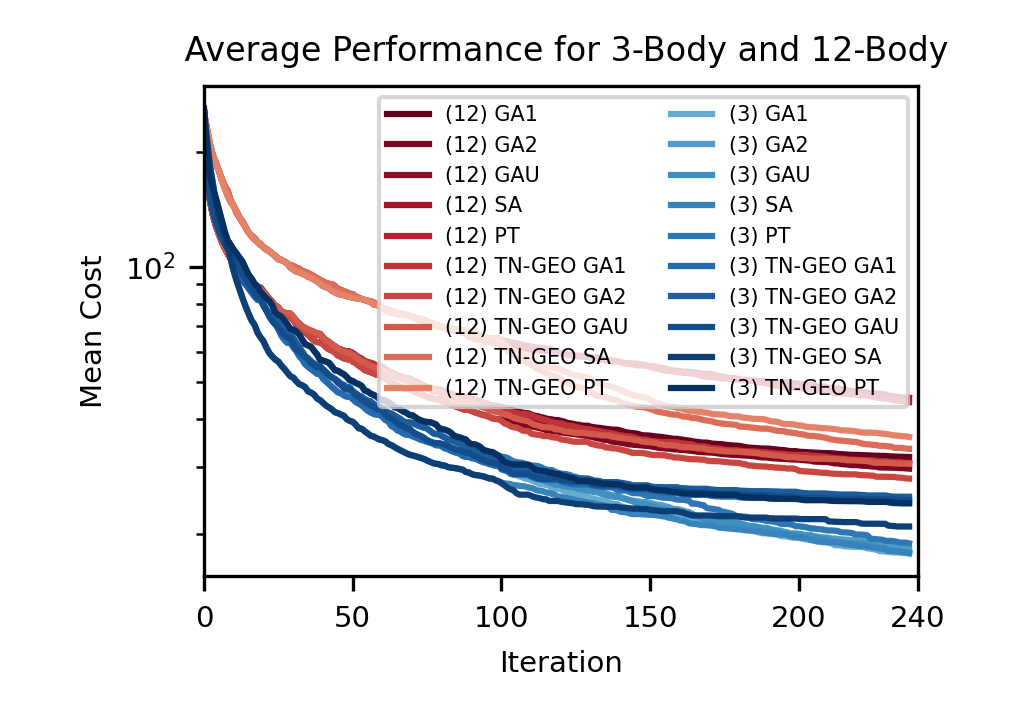}
        \caption{Performance of all optimizers for 3-body and 12-body formulations of the problem. We note that $0$ as defined in this plot is not the global minimum but instead is the global minimum of a simplified form of the problem, as the true global minimum is not known. See text for the detailed protocol of calculations.}
        \label{fig:configdif}
        \end{figure}

        Our first numerical investigation explored the difference in optimizer performance owing to problem parameterization. Each optimizer is implemented as described in Appendix \ref{sec:traditional-solvers} and Section \ref{sec:methods of opt} and allowed to run for an empirically-based choice of 240 cost-function evaluations (TN-GEO) being performed explicitly on only the final 140, as described in Section \ref{sec:methods of opt}. These sets of optimization trials were then repeated using 300 randomized initial states to acquire statistical data. The averaged results of this study are plotted in \figref{fig:configdif}. All optimizers, regardless of their type (conventional or quantum-inspired) perform significantly better when the problem is formulated in the 3-body parameterization, achieving lower costs in fewer iterations on average.
        %%%%%%%% FIGURE: parameterization comparison %%%%%%%%
        
        % Following this investigation, we considered the performance of the RBM-augmented optimizer as compared to the best-performing conventional optimizers in the realistic setting where the problem space is reduced due to clever preprocessing. The results of this reduction are given in \ref{table: performance over reduction} where we denote the optimizers that achieve the lowest minima, as well as the best-performing optimizers on-average.

       %%This parameterization is found to be suboptimal (see Section~\ref{sec:numerical_results}) for two reasons: The first is that the parameterization naturally places inter-stage and intra-stage interactions on the same level, which makes little intuitive sense when same-stage shops have similarly achievable production levels. In addition, the simple but unintuitive parameterization makes it difficult to identify or compactly search relevant subspaces.
        
        Following this investigation, we focused solely on the 3-body parameterization and performed a comparison of binary encoding methods across all state spaces given in Table \ref{tab:solution_space_size} (basic, Gray and PGGray encodings). 
        As before, for each scenario, we ran each conventional solver 300 times to accumulate statistics, performing up to 240 cost evaluations each time. A description of the conventional solvers and hyperparameters used in these benchmarks are provided in Appendix~\ref{sec:traditional-solvers}. 

        For TN-GEO an important hyperparameter is the \emph{maximum bond dimension} of the Matrix Product State which controls the expressivity of the tensor network and is an indicator of the correlation radius in the training data. For the problem at hand, we tested TN-GEO performance while sweeping this parameter, as shown in \ref{sec:bond_dimension_optimization}. Through this process we found that a maximum bond dimension of 6 is optimal as it achieves equal or better performance (across solvers and state space sizes) as compared to other bond-dimensions.

        \begin{figure}
            \centering
            \includegraphics[width=0.95\columnwidth]{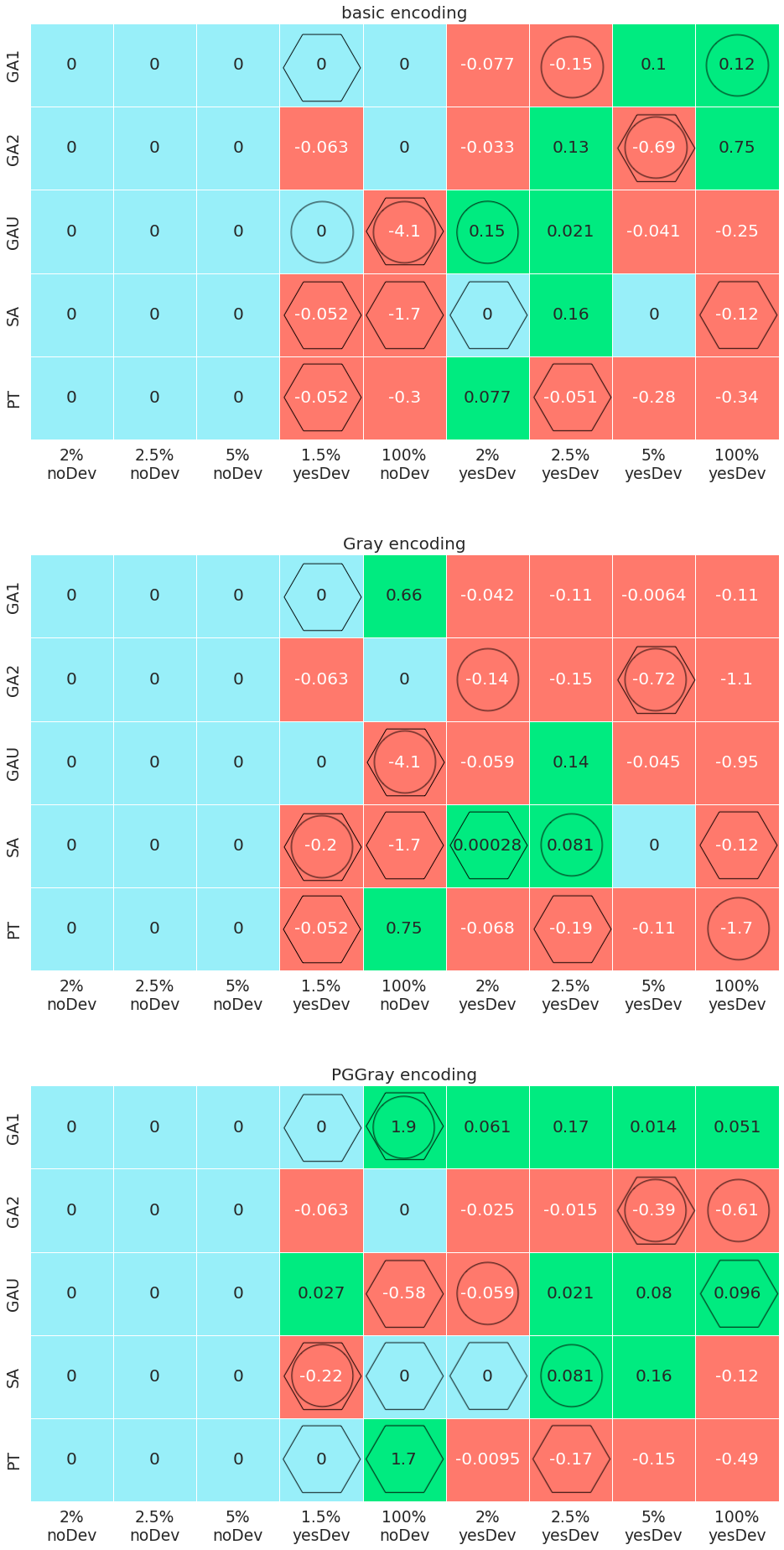}
            \caption{Performance of TN-GEO in 3-body formulation. This figure contains one heatmap for each state encoding, each depicting the difference between the lowest cost found by a conventional solver (one of GA1, GA2, GAU, SA or PT) after 300 independent runs, and the lowest cost found by TN-GEO boosting the same conventional solver for the same 300 runs, in different problem configurations (with/without deviation and different margins). The difference is positive (green cells) when the best cost found by TN-GEO is strictly smaller than the best cost found by the conventional solver, zero (blue cells) when they tie in their best run, and negative (red cells) when TN-GEO could not achieve the lowest cost found by the conventional solver. For each problem configuration, there is (at least one) cell with a circle/hexagon indicating that either the corresponding conventional solver or TN-GEO boosting that conventional solver is the worst/best solver among all of the considered solvers. In the event of a tie, multiple circles and hexagons appear.}
        \label{fig:GEO-vs-conventional-solvers-3body}
        \end{figure}
        
        \begin{figure}
            \centering
            \includegraphics[width=0.9\columnwidth]{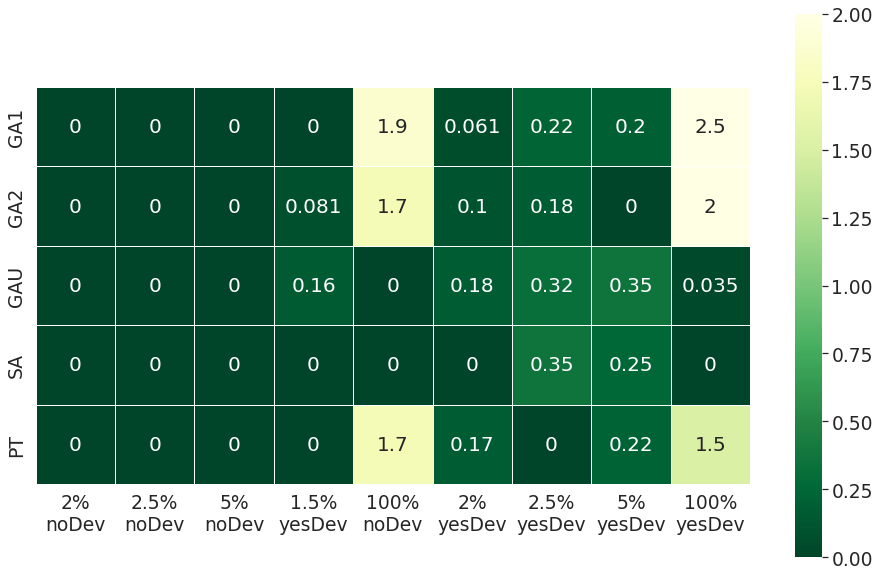}
            \caption{Relative performance of the conventional solvers. Each cell in the grid gives the difference between the minimum achieved by the best conventional solver and the minimum achieved by the conventional solver on the left axis. The overlaid heatmap allows for quick comparison of these values with green indicating that the solver performed nearly or exactly as well as the best solver.}
            \label{fig:conventional_min}
        \end{figure}

        \begin{figure}
            \centering
            \includegraphics[width=0.9\columnwidth]{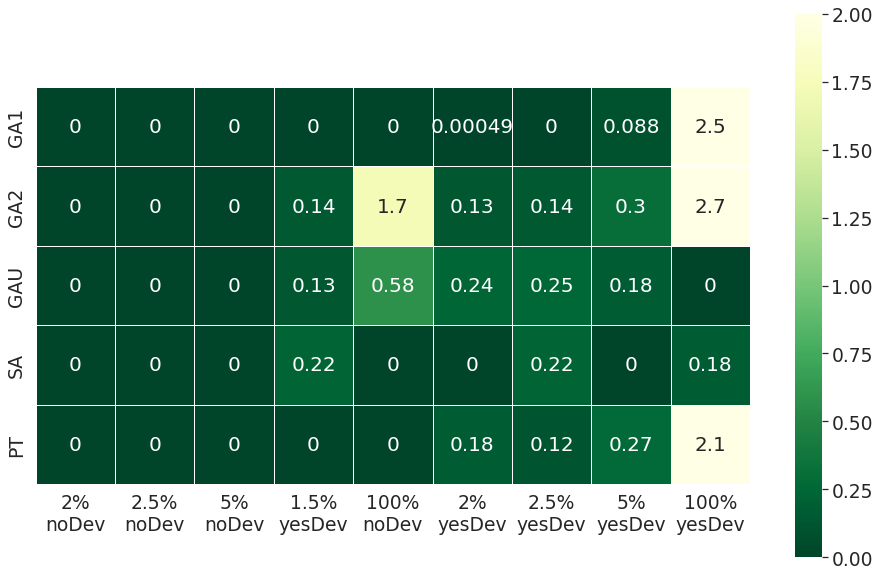}
            \caption{Relative performance of TN-GEO trained with data obtained from the conventional solvers. Each cell in the grid shows the difference between the minimum achieved by the best TN-GEO solver and the minimum achieved by TN-GEO boosting the solver on the left axis. This is distinct from \figref{fig:conventional_min} which only compares the conventional solvers with the best conventional solver.}
            \label{fig:geo_min}
        \end{figure}
        
        The results of these studies (for a maximum bond dimension of 6) are depicted in Fig.~\ref{fig:GEO-vs-conventional-solvers-3body}, where there is a heatmap for each state encoding. The number within each heatmap square indicates the largest amount (of the 300 runs) by which TN-GEO, when trained using the first 100 cost function evaluations of a traditional solver, can outperform the the conventional solver used for training. To facilitate a more detailed analysis Fig.~\ref{fig:conventional_min} shows the performance of the conventional solvers with respect to each other and Fig.~\ref{fig:geo_min} depicts the dependence of the relative performance of TN-GEO using the PGGray encoding with respect to the conventional solver results used as the training set.

            % Comparing the heatmaps in Fig.~\ref{fig:GEO-vs-conventional-solvers-3body}, one can see that using a problem-inspired encoding, i.e., PGGray encoding, helps TN-GEO to achieve better results. In particular, using PGGray encoding TN-GEO could improve the performance of conventional solvers in more than 71\% of possible scenarios. 
    \section{Discussion}
    
        Before we take a closer look at the difference between various solvers, let us note that the 3 smallest cases (2\% noDev, 2.5\% noDev, 5\% noDev) are so small that all solvers reached exactly the same minimum (see Figs.~\ref{fig:conventional_min} and \ref{fig:geo_min}), which we verified to be the global minima for those cases. Hence, to differentiate between the solvers we use the results of the remaining (larger) state space cases only.

        When considering the performance of the conventional solvers, SA performs the best across the board. However, when we additionally remove the cases in which the solution space is not reduced (100\% noDev, 100\% yesDev), GA2 becomes the best solver, closely followed by PT. Excluding these cases  from our conclusions is justified as these cases are not used in practical applications and are apparent outliers.
        
        The issue of TN-GEO performance is more complicated. Firstly, the encoding of the states plays a significant role. The PGGray encoding is the clear best, as demonstrated in Fig.~\ref{fig:GEO-vs-conventional-solvers-3body}. When considering only this best encoding, and again disregarding the non-reduced cases, we can see in Fig.~\ref{fig:geo_min} that TN-GEO trained on the data obtained from GA1 performs best of all TN-GEO variants. Given the relatively poor results of GA1 with respect to the best conventional solvers, it becomes apparent that the quality of a generated training set is not directly correlated with the performance of the conventional solver.

    \section{Summary}
    
        We investigated the utility of quantum-inspired, generator enhanced optimization methods in a realistic use-case: an instance of a BMW production planning problem. Through this comparison we found that problem representation plays a significant role, with parameter groupings based on known correlations allowing for lower minima to be achieved by all solvers more quickly. In addition, we found that problem-motivated encoding of the assembly line parameters leads to greatly improved results as compared to na\"ive representations. Combining these two techniques, we show that quantum-inspired generative model based solvers tie or improve upon the tested conventional optimization methods in a majority of the tested cases, particularly when used in intermediate solution space sizes. Based on this comparison, we conclude that TN-GEO ties or outperforms respective tested conventional optimization methods in 31 out of 45 tested cases, demonstrating the power of these optimization methods in industrially-relevant settings.  
    \section{Statement on Data Availability}
        \label{sec:data_availability}
        %Source code of all the conventional solvers as well as additional tools for data analysis and all acquired data is hosted at 
        All data used in this paper is publicly available at \url{https://github.com/zapatacomputing/TN-GEO-car-plant-optimization}.% this will be moved over to a publicly accessible repository
    \section{Acknowledgements}
    
         The authors would like to thank the Zapata Quantum Machine Learning team for their helpful insight into the data analysis and feedback on the manuscript, in particular A. Perdomo-Ortiz and M. Mauri. This work was funded in part by the NTT Phi Laboratory and by BMW and Zapata Computing through the Quantum Science and Engineering Consortium (QSEC) at the MIT Center for Quantum Engineering (CQE). The authors have declared that no competing interests exist.
    
    \FloatBarrier
    
    %\section*{References}
    \bibliography{main.bib}

    \appendix

    \section{The free-stage approximation}

    \subsection{Production estimation}
    \label{sec:production_estimation}
    
    In the actual model, the production cost is determined both by the state of every stage and the inter-stage interactions. In order to simplify the problem, we introduce the free-stage approximation, in which we neglect the inter-stage interactions. At this level of approximation, we can decouple the model and consider every stage separately.
    
    We will start by introducing \(\tilde{p}_{n,m}\) as the estimated monthly production of stage \(n\in\{1,\ldots,N\}\) being in the state \(m\in\{1,\ldots,M\}\). This quantity can be readily calculated by taking the worktime of the stage obtained from its state, multiplying it by the production rate of the shops comprising the stage, and summing the results. Obviously, for every configuration \(m_1,m_2,\ldots,m_N\)
    \[
    \begin{split}
        p_{m_1,m_2,\ldots,m_N} & \leqslant\tilde{p}_{m_1,m_2,\ldots,m_N} = \\
        & \min\{\tilde{p}_{1,m_1}, \tilde{p}_{2,m_2},\ldots,\tilde{p}_{N,m_N} \}
    \end{split}
    \]
    where \(p_{m_1,m_2,\ldots,m_N}\) and \(\tilde{p}_{m_1,m_2,\ldots,m_N}\) is the actual and approximate monthly production, respectively. Additionally, we can expect that
    \[
        \frac{|p_{m_1,m_2,\ldots,m_N} - \tilde{p}_{m_1,m_2,\ldots,m_N}|}{\tilde{p}_{m_1,m_2,\ldots,m_N}} < \varepsilon
    \]
    for a relatively small value of \(\varepsilon\), although the concrete value will depend on the specific parameterization of the model and is difficult to predict accurately. Numerical experiments suggest the \(\varepsilon\) value for the global minimum configuration between about \(0.01\) and \(0.025\), depending on the parameterization.
    
    \subsection{Reduction of the solution space}
    \label{sec:reduction}
    For this reason, it should be sufficient to restrict the search for the global minimum to the reduced search space
    \[
    \mathcal{R}=\left\{(m_1,m_2,\ldots,m_N)\in\mathcal{S}|1-\varepsilon\leqslant\frac{\tilde{p}_{i,m_i}}{p_t}\leqslant1+\varepsilon)\right\}
    \]
    using the value of of \(\varepsilon\) estimated above. We note that the cost of the reduction scales proportionally to the number of single-stage states, thus avoiding the exponential explosion.
    
    \subsection{The production-guided cost-optimization algorithm}

    The approach presented above can be extended, allowing us to formulate an efficient method of exploring the solution space, the Production Guided Cost Optimization (PGCO) algorithm.
    
    The cost function penalizes not meeting the target production, which we have taken advantage of above, as well as under-utilization of the production stages. Hence, for the optimal configuration \((m_1,m_2,\ldots,m_N)\) we should expect
    \[
        p_t\approx\tilde{p}_{1,m_1}\approx\tilde{p}_{2,m_2}\approx\ldots\approx\tilde{p}_{N,m_N}.
    \]
    
    This observation allows us to easily formulate an efficient algorithm for exploring the solution space. Namely, we construct a forest of single stage-states. First, we pick as roots \(N_r\) first-stage states for which estimated production is closest to the target. Then, for each root we pick \(N_b\) second-stage states for which the estimated production is closest to the estimated production of the respective root. Then we repeat the procedure, analogously extending every branch of every tree. Finally, we exhaustively search for the minimum among the generated \(N_rN_b^{N-1}\) states.
    
    The weak side of the approach is calibration, required to determine values of \(N_r\) and \(N_b\) as small as possible, but at the same time, large enough to ensure the presence of a global minimum in the spanned subspace.
    
    On the other hand, assuming relatively large values of \(N_r\) and \(N_b\), PGCO can be considered as a preprocessing stage, efficiently reducing the search space for a generic global minimization procedure.
    
    \subsection{Production guided encoding}
    \label{sec:pg_encoding}
    
    The GEO optimizer requires all states of the optimized system to be represented as bitstrings (sequences of binary digits), In principle, we can just enumerate the states and represent them as binary numbers, padding with zeros when necessary. In our case, the simplest method boils down to representing the state as an \(N\)-digit base-\(M\) number and transforming it to base 2. Unfortunately, such a mapping may distort the topology of the solution space, thus making training the generative model much more difficult.
    
    The first source of such distortion is the standard representation of non-binary numbers in base 2, which causes certain neighbouring numbers, like 3 and 4, to be represented by bit-strings having relatively large Hamming distance to each other. Fortunately, this can be mitigated by using the Gray encoding.
    
    The other significant factor impacting the effectiveness of the representation is the original representation of system states as \(N\)-digit base-\(M\) numbers. In this approach nearest neighbours may have significantly different production cost. To alleviate the issue we came up with the Production-Guided Encoding of the states. In this approach, we enumerate single-stage states of the first stage according to the absolute value of the difference between the estimated production of the first stage and the target production. This way, we obtain indexing of the first stage single-stage states, where the lower is the index, the closer to the target is the production. Then, for every single-stage state from the first stage, we enumerate the second stage single-stage states, ordering them by how similar their production is to that of the considered first-stage state. Finally, for every pair of the first and second stage single-stage states, we enumerate the third stage single-stage states, ordering them by how similar is their production to that of the considered first stage state. This way we dynamically build a three-level forest, in which values the indices of the branches approximately represent the deviation from the expected target production. Those branch indices are in turn represented by bit-strings using the Gray encoding.
    
    \section{Traditional solvers}
    \label{sec:traditional-solvers}
    
    In this work, we used Genetic Algorithms, Simulated Annealing, and Parallel Tempering solvers both to generate initial learning data for TN-GEO and as benchmarks. All optimization methods were implemented in python 3.8. Below we provide brief description of the specific implementations we used.

    \subsection{Genetic Algorithms}
    The genetic algorithm is a  meta-heuristic algorithm inspired by the theory of natural evolution~\cite{holland1992:adaptation_natural}. For implementing the genetic algorithm, we used the DEAP~\cite{FDGPG2012-DEAP} library equipped with elitism mechanism from Ref.~\cite{wirsansky2020:hands-on_genetic}. We initialized the solver with a population of 10 individuals. In each iteration, a subset of size 9 of the population from the previous iteration is selected using tournament among 3 individuals. Then, a new population of the same size is generated \textit{via} mutation (with probability 0.8) of each individual in the subset and crossover (with probability 0.8) of consecutive ones. At the end of an iteration, the best individual observed in all previous iterations is added to the population. We used three different types of crossover operations: one-point, two-point, and uniform. For mutation, one shop (or one stage in 3-body formulation) is chosen uniformly at random, and its state is updated to a random possible state.

    \subsection{Simulated Annealing}
    For Simulated Annealing (SA), we implement a version of the algorithm based on Ref~\cite{KGV1983-SA}. In our implementation, starting from an initial state with cost \( C_0\), the initial temperature \( T_0\) is set to 50, and in each iteration, the state of one shop (or one stage in 3-body formulation), chosen uniformly at random, is updated to a random state with probability 
    \[
    p_i = \min\{1, \exp((C_{i-1} - C_{i})/T_{i-1})\} 
    \]
    where \(C_i\) is the cost of the state at the end of the \(i\)-th iteration. At the end of each iteration, the temperature is decreased by a factor of 1.2.
    
    \subsection{Parallel Tempering}
    For Parallel Tempering (PT), we implemented a version of the algorithm based on Ref.~\cite{HANSMANN1997-PT}. PT can be thought of as an advanced version of Simulated Annealing, where one considers independent copies of the car plant's state, called replicas, each at a different temperature \(T_{r}\). In each iteration, the state of each replica is updated according to its temperature, as in the SA algorithm. Then, the state of two neighbouring replicas, \( r\) and \(r+1\), are swapped with probability 
    \[
    p_r = \min\{1,\exp((C_r-C_{r+1})(\beta_{r}-\beta_{r+1})\} 
    \] 
    where \( C_{r}\) is the cost of the state in \(r\)-th replica, and \( \beta_r = 1/T_{r}\). In our implementation, we had 5 replicas with 4 state-updates in each iteration (before each replica swap), and we used a geometric sequence of evenly spaced (in log space) numbers between 0.1 and 10 for \( \beta_r \) values.

    \section{TN-GEO Bond-Dimension}
    \label{sec:bond_dimension_optimization}
    
    \begin{figure}
        \centering
        \includegraphics[scale = 1.0]{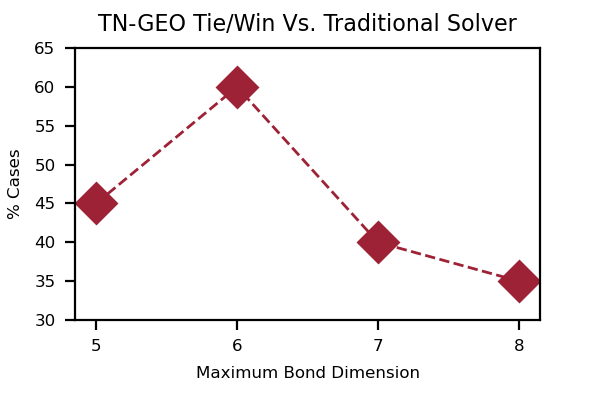}
        \caption{The percentage of total tested cases (combinations of margin, rate deviation and traditional optimizer) in which TN-GEO improves upon the performance of the respective traditional optimizer. A maximum of cases occurs when a maximum bond dimension of 6 is used for the TN-GEO matrix product state optimizer.}
        \label{fig:bond_dim_opt}
    \end{figure}

    When using TN-GEO, a fixed hyperparameter, the maximum bond dimension, sets the maximum number of singular values kept in the matrix product state factorization of correlation tensors. This corresponds to the level of correlation that can be captured by the model. Too high a maximum bond dimension is likely to overfit a given dataset, while too low does not capture relevant features.
    
    We tested TN-GEO directly on the studied optimization problem in order to choose the maximum bond dimension that provides best performance. For each bond dimension, we tabulated the number of cases (tested combinations of margins, rate deviation and traditional optimizer) in which the TN-GEO optimizer improves upon the performance of the respective traditional optimizer. We anticipated a ``sweet-spot" of performance would exist across maximum bond dimensions, and found one at a maximum bond-dimension of 6 singular values (see \figref{fig:bond_dim_opt}) when using 3-body formulation and the production-guided encoding. 

    \section{Time Domain Simulation}
        \label{sec:TD Sim}
        The time domain simulation of the assembly line was implemented in python and was a modified form of that used in BMW manufacturing settings. The simulation was performed at a half-hour timestep to maintain accuracy while minimizing runtime. Over the course of the optimization 81\% to 69\% of the runtime was consumed by this simulation.
        % will need to estimate
    
\end{document}